\documentclass[journal]{IEEEtran}

\usepackage{amsmath,amsfonts}
\usepackage{graphicx}
\usepackage{booktabs}
\usepackage{cite}
\usepackage{xcolor}
\usepackage[hidelinks]{hyperref}

\hypersetup{
    pdftitle={Reducing Memory Traffic in FP4 LLM Training},
    pdfauthor={Chandra Sekhar Mummidi, Sandip Kundu},
    pdfsubject={GPU Architecture and Low-Precision Training},
    pdfkeywords={FP4 Training, NVFP4, MXFP4,
                 Large Language Models, Low-Precision Training,
                 GPU Architecture, NVIDIA Blackwell,
                 Transformer Engine, Kernel Fusion,
                 Gradient Quantization, Gradient Accumulation,
                 Memory Traffic}
}

\definecolor{oldtext}{RGB}{163,74,74}
\definecolor{newtext}{RGB}{0,102,153}

\begin{document}

\title{Reducing Memory Traffic in FP4 LLM Training}

% \author{Chandra Sekhar Mummidi and Sandip Kundu%
% \IEEEauthorblockA{
% University of Massachusetts Amherst, MA, USA\\
% \texttt{\{cmummidi, kundu\}@umass.edu} 
% \vspace{-10mm}
% }

\author{
{\large Chandra Sekhar Mummidi and Sandip Kundu}\\[0.5em]
University of Massachusetts Amherst, MA, USA\\
\texttt{\{cmummidi, kundu\}@umass.edu}
\vspace{-10mm}
}

\maketitle

\begin{abstract}
Native FP4 training retains higher-precision gradient computation and weight-gradient accumulation. Our characterization of NVIDIA Transformer Engine on Blackwell identifies two sources of intermediate global-memory traffic: a numerical boundary, where unavailable scaling information forces gradient materialization, and a kernel boundary, where separate microbatch GEMMs repeatedly store and reload FP32 accumulators. Together, these create a global-memory boundary not inherently required by FP4 arithmetic. We address the numerical boundary through \emph{producer-local FP4 quantization}, converting gradients within their producing kernels to avoid intermediate BF16 transfers. We address the kernel boundary through register-resident FP32 accumulation across microbatch GEMMs, preserving accumulation order. Under a fixed operand-memory budget, two-microbatch windows across all projections recover 69\% of the training-step time savings of four-microbatch accumulation across all projections with roughly one-third of the additional operand storage. On Blackwell, our implementation reduces GPU kernel time by 9.28\% and improves CUDA Graph training throughput by 9.63\%, with a sustained 9.7\% improvement over 20 million training tokens and comparable validation perplexity to Transformer Engine NVFP4. FP32 accumulation matches the sequential reference bit for bit across more than 1.2 billion elements. On AMD MI210, optimized software-emulated MXFP4 execution using BF16 GEMMs improves training throughput by 30.9\% over the corresponding MXFP4 reference.

\end{abstract}


\section{Introduction}
\label{sec:introduction}

Training large neural networks involves repeated matrix multiplications, so efficient general matrix multiplication (GEMM) implementation is key to reducing overall training cost \cite{Panferov2026QuartetIA}. Moving to FP4 can reduce this cost, but the narrow range and limited precision of four-bit values make them challenging to use in training. Recent FP4 training methods have addressed these numerical difficulties. Fine-grained scaling helps map values of different magnitudes into the limited FP4 representable range.  Stochastic rounding reduces rounding bias, while randomized transforms redistribute large values to reduce their impact on quantization. Together, these techniques have enabled greater use of FP4 operations in large language model (LLM) training ~\cite{Tseng2025TrainingLW,Castro2025QuartetNF,Panferov2026QuartetIA,Cook2025FourOS,Chen2025TetraJetv2AN,Nvidia2025PretrainingLL,Chmiel2025FP4AT}. NVIDIA Blackwell provides native Tensor Core support for block-scaled FP4 matrix multiplication, allowing these low-precision computations to execute directly on specialized hardware. Yet much of the backward computation surrounding these matrix multiplications still operates at higher precision. In the Blackwell backward path we examine, preparing FP4 gradients and accumulating weight gradients involve intermediate global-memory accesses that contribute to execution time.

Native FP4 Tensor Core execution does not eliminate higher-precision intermediate values in the backward pass.
Our characterization of the NVIDIA Transformer Engine backward path on Blackwell reveals two sources of intermediate global-memory traffic. The first arises from a \emph{numerical boundary} associated with FP4 scale availability: activation gradients are written to global memory in BF16 and read by the preceding layer for FP4 conversion, even though the GEMMs producing them use
FP4 operands. The second arises from a \emph{kernel boundary} during weight-gradient accumulation: separate
microbatch GEMMs repeatedly store and reload the FP32
accumulator. Together, these boundaries create a
\emph{global-memory boundary}, where intermediate
gradients and accumulators are transferred through global
memory. Transformer Engine already fuses scaling,
encoding, and operand-layout preparation within its
precision conversion kernels
~\cite{transformerengine,NVIDIA2026TransformerEngineNVFP4},
but these optimizations do not eliminate the identified
intermediate transfers. This raises an important
architectural question: \textit{which global-memory
transfers are required by FP4 numerical requirements,
and which arise from how backward computation is
organized into GPU kernels}?

The numerical and kernel boundaries do not inherently require intermediate global-memory transfers. For the
numerical boundary, making the required FP4 scale available during gradient computation allows the gradient to be
converted directly to FP4 \emph{within the gradient-producing kernel}, without first materializing it in BF16. For the kernel boundary, weight-gradient GEMMs can be deferred until their operands are available~\cite{Qi2024ZeroB,NVIDIA2026MegatronCore}, since weights are updated only after gradient accumulation. Successive microbatch GEMMs can then accumulate into the same FP32 register tile in the original microbatch order, avoiding intermediate accumulator transfers.

Avoiding intermediate FP32 accumulator transfers by deferring weight-gradient GEMMs requires retaining their FP4 inputs in GPU memory until they execute. Retaining the inputs from every microbatch can consume substantial memory, particularly when several layers and projections use deferred accumulation. Rather than waiting for all microbatches, we can execute the weight-gradient GEMMs in smaller groups. We refer to the number of microbatches accumulated together as the \textit{accumulation window}. A larger window reduces repeated FP32 accumulator accesses but requires more memory to retain the FP4 inputs. It can also change GEMM execution efficiency, so a larger window does not necessarily give the shortest execution time. The choice of window size and the layers and projections that use deferred accumulation must therefore account for both execution time and available GPU memory.

Prior FP4 training systems have addressed parts of these execution costs. CUTLASS supports FP4 conversion in GEMM kernels, while Quartet-II and Format-Aware Fusion combine gradient-related computations with FP4 encoding~\cite{cutlass,Panferov2026QuartetIA,Hu2026FormatAwareFF}. Transformer Engine and Megatron-LM accumulate weight gradients in FP32, storing the results in global memory between microbatches~\cite{transformerengine,NVIDIA2026MegatronAccumulation}. FORGE combines weight-gradient computation with optimizer updates, while Zero Bubble defers weight-gradient computation to improve pipeline scheduling~\cite{Kukreja2026FORGEFO,Qi2024ZeroB}. These methods address different training costs but do not eliminate intermediate global FP32 accumulator accesses across separately executed microbatch GEMMs in backpropagation.

We develop a GPU execution scheme that reduces
global-memory accesses during FP4 gradient preparation
and weight-gradient accumulation. Our implementation
reduces GPU kernel time by 9.28\% and improves
Llama 3.2 3B training throughput by 9.63\% under
CUDA Graph execution. Across more than 1.2 billion
weight-gradient elements, the FP32 accumulation
results match the sequential reference bit for bit.

Under the same additional memory budget, shorter
accumulation windows distributed across more projections
can outperform longer windows restricted to fewer
projections. Thus, the benefit of register accumulation
depends on how the available memory is distributed
across projections, not simply on the number of
microbatches accumulated together. Accumulating all
four microbatches across all projections yields the
lowest step time in our experiments but exceeds
available GPU memory at the longer sequence length
we tested. Over 20 million tokens of continued
pretraining, the implementation sustains approximately
10\% higher throughput than Transformer Engine NVFP4,
and achieves a validation perplexity of 11.053,
compared with 11.117 for the reference.

Our contributions include:

\begin{itemize}

\item Identification of the numerical boundary in NVFP4
gradient preparation and development of producer-local
FP4 conversion using causal scaling.

\item Development of register-resident FP32 accumulation
across microbatch GEMMs, preserving the original
accumulation order without intermediate accumulator
transfers.

\item Characterization of accumulation-window selection
under operand-memory constraints, balancing memory
traffic, execution time, and projection coverage.

\item Evaluation on NVIDIA Blackwell and AMD MI210,
demonstrating 9.7\% and 30.9\% higher training
throughput over their respective FP4 references.

\end{itemize}

\section{Background and Prior Work}
\label{sec:background}

\subsection{Scaling in FP4 Training}
\label{sec:background_fp4}

FP4's limited numerical range makes scaling important during training. Blackwell supports the E2M1 FP4 format, which uses two exponent bits, one mantissa bit, plus a sign bit, with a maximum finite magnitude of 6 and a smallest positive value of 0.5. Values outside this range cannot be represented directly in finite FP4. Values with an absolute magnitude greater than 6 are clipped to that limit. At the other end, values with magnitude smaller than 0.5 may round to zero. Scaling adjusts the magnitude of the input before quantization and restores it afterward. For an input $x_i$ and a positive scale $s$, the quantized value is reconstructed as

\begin{equation}
    \hat{x}_i = s\,Q_{\mathrm{E2M1}}\left(\frac{x_i}{s}\right),
    \label{eq:fp4_scaling}
\end{equation}

Here, $x_i$ is the original input value, $s$ is the scaling factor, and $Q_{\mathrm{E2M1}}$ rounds a value to the E2M1 format. The input is first divided by $s$ to bring it into the FP4 range. After quantization, the result is multiplied by $s$ to obtain the approximate original value $\hat{x}_i$. A common choice is $s = \max_i |x_i|/6$, which maps the largest absolute input value to the maximum E2M1 magnitude of 6 before rounding.

Scaling has long been used in mixed-precision training. NVIDIA's automatic mixed precision (AMP) supports loss scaling to prevent small gradients from underflowing during FP16 backpropagation~\cite{Micikevicius2017MixedPT}. BF16 has a wider exponent range than FP16 and generally does not require the same loss-scaling treatment.

FP4 operand scaling serves a different purpose: it selects the numerical range used to encode matrix operands. A single scale for an entire tensor is inexpensive to store. However, for a narrow FP4 range, it is difficult to represent a tensor whose values span several orders of magnitude with a single scale. Large values in the tensor can determine the quantization resolution for all other values. Finer-grained scaling assigns different scales to smaller groups of elements in a tensor, allowing their quantization ranges to follow local magnitudes.

MXFP4 uses this approach with an E8M0 (eight exponent bits and no mantissa bits ) power-of-two scale shared by 32 E2M1 values~\cite{Tseng2025TrainingLW,Rouhani2023MicroscalingDF}. Each block can select its scale independently, avoiding the need for a tensor-wide maximum. However, adjacent E8M0 scale values differ by a factor of two, restricting how finely the scale can be adjusted. NVFP4 uses a different hierarchy on NVIDIA Blackwell: each group of 16 E2M1 values has an E4M3 block scale, together with an FP32 scale shared across the tensor~\cite{Nvidia2025PretrainingLL,NVIDIA2026TransformerEngineNVFP4}. The reconstructed value follows

\begin{equation}
    \hat{x}_i = S\,s_{b(i)}\,Q_{\mathrm{E2M1}}\left(\frac{x_i}{S\,s_{b(i)}}\right),
    \label{eq:nvfp4_scaling}
\end{equation}

where $S$ is the tensor-level FP32 scale and $s_{b(i)}$ is the E4M3 scale associated with the block containing $x_i$. E4M3 provides finer scale choices than the power-of-two E8M0 format, while the FP32 global scale allows the block scales to represent magnitudes beyond their individual numerical range.

The choice of scaling granularity also affects GPU execution. In MXFP4, each scale depends on a group of 32 values and can be computed locally without synchronization across thread blocks. NVFP4 also uses an FP32 tensor-level scale. When this scale is calculated from the gradient being computed, each thread block first computes the maximum absolute value of its portion of the gradient. These local maxima must then be combined to find the maximum across the entire tensor~\cite{Panferov2026QuartetIA,Hu2026FormatAwareFF,NVIDIA2026TransformerEngineNVFP4}. FP4 conversion must wait for this reduction to finish, even when individual thread blocks have already computed their gradient values.

Delayed scaling, previously used in FP8 training, provides a scaling factor from earlier tensor statistics~\cite{transformerengine,Fishman2024ScalingFT}. Historical gradient scaling has also been explored for FP4 training~\cite{Sun2020UltraLowP4}, while recent UE5M3 work uses periodic tensor scaling~\cite{Hu2026UE5M3FB}. Applying this approach to the FP32 global scale in NVFP4 allows the scale to be available before gradient computation, while the E4M3 block scales can still be computed from the current gradient. 
Flexpoint uses historical statistics for scale
prediction~\cite{Kster2017FlexpointAA}, while MOSS
predicts weight scales to avoid runtime
reductions~\cite{Zhang2025MOSSEA}.

\begin{figure*}[t]
    \centering
    \includegraphics[width=0.7\textwidth]
        {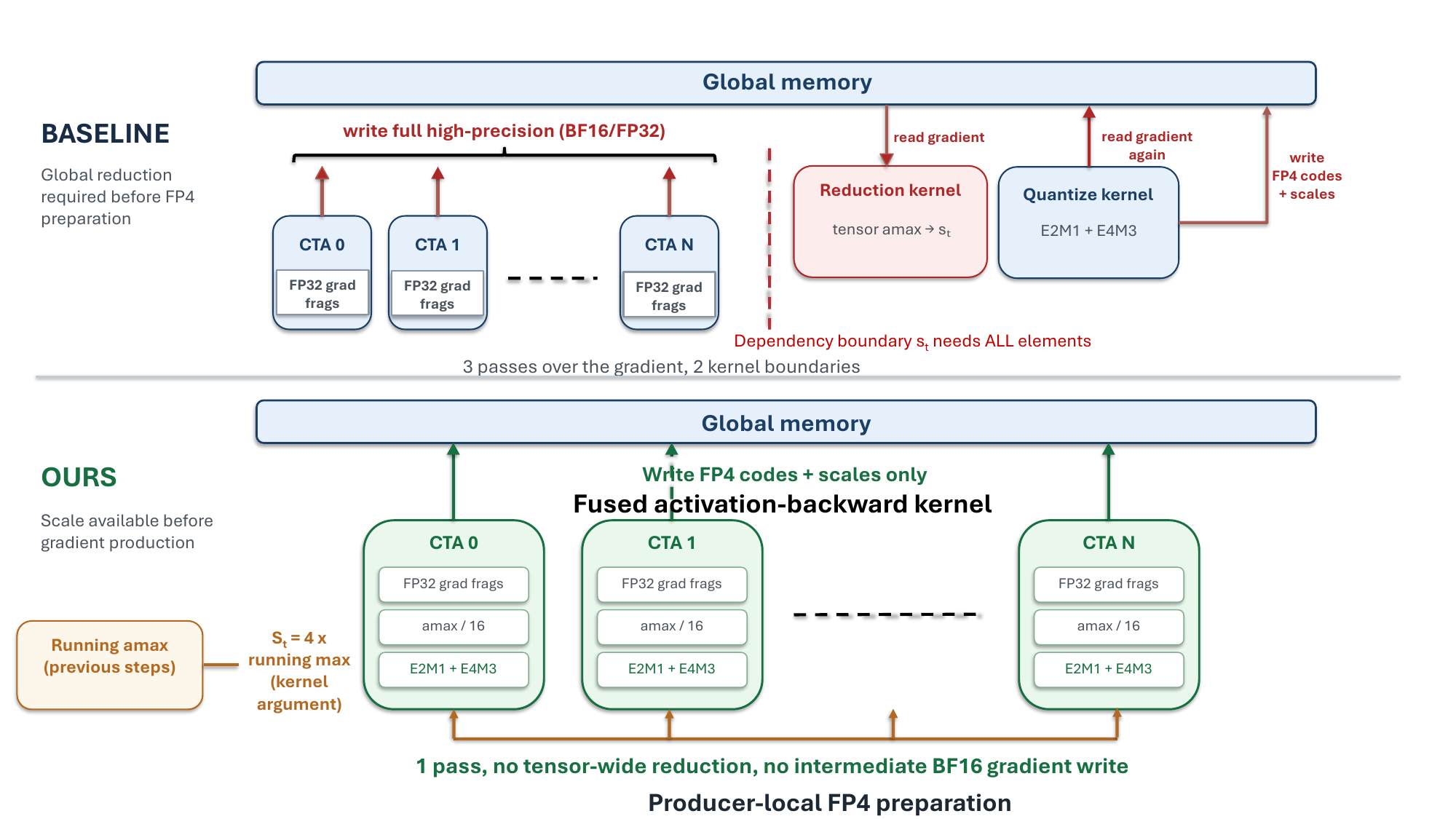}
    \caption{\textbf{Numerical boundary:} Producer-local FP4 gradient quantization compared with a simplified Transformer Engine preparation path on Blackwell SM120. CTA denotes Cooperative Thread Array (CUDA thread block).}
    \label{fig:numerical_boundary}
\end{figure*}
\subsection{Gradient Conversion and Fusion}
\label{sec:background_quantization}

Figure~\ref{fig:numerical_boundary} illustrates the numerical boundary in Transformer Engine's FP4 gradient preparation~\cite{transformerengine}. On Blackwell SM120, the gradient is first written to global memory in BF16. A separate kernel computes the tensor-wide maximum magnitudes, followed by a quantization kernel that reads the gradient to generate the rowwise FP4 operand. A randomized Hadamard transform reads the gradient again and writes a BF16 intermediate, which another kernel converts into the columnwise FP4 operand. Separate kernels prepare the block-scale layouts required by subsequent GEMMs. The figure simplifies these operations into a common FP4 preparation stage; the actual implementation uses separate kernels for rowwise and columnwise preparation.

\begin{figure*}[t]
    \centering
    \includegraphics[width=0.7\textwidth]
        {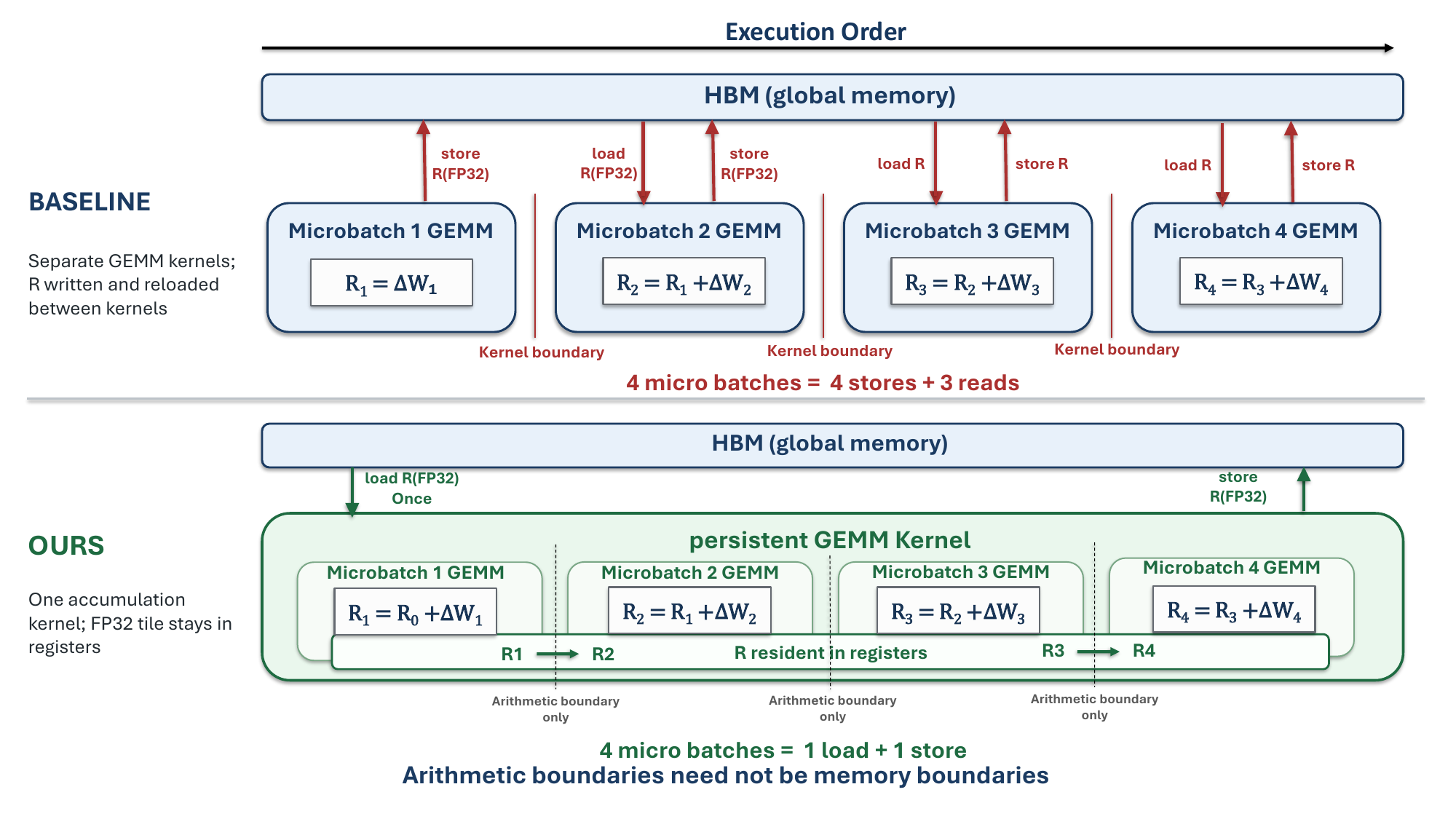}
    \caption{\textbf{Kernel boundary}
    Register-resident FP32 accumulation across microbatch GEMMs.}
    \label{fig:kernel_boundary}
\end{figure*}

One approach to reducing FP4 preparation overhead is to perform conversion within the GEMM itself. CUTLASS supports converting accumulator values directly to FP4 in the GEMM epilogue~\cite{NEW_cutlass-blackwell-docs}. 
FP4 training on Hopper GPUs has also used direct
precision conversion and scaling-aware operand-layout
preparation~\cite{Zhang2026PracticalFT}. More broadly, CODA fuses Transformer operations into
GEMM epilogues to reduce intermediate global-memory
materialization~\cite{Guo2026CODART}. Another approach is to combine multiple preparation operations in a single kernel. Hadamard transformations are used before quantization to redistribute large values across a group, reducing the effect of outliers on FP4 quantization. 

Quartet-II combines this transformation with FP4 encoding,
while Format-Aware Fusion produces the rowwise and
columnwise FP4 operand layouts required by backward
GEMMs~\cite{Panferov2026QuartetIA,Hu2026FormatAwareFF}.
 
However, when preparation begins from a gradient already stored in global memory, combining these operations does not eliminate the intermediate gradient write and read.

Our implementation combines FP4 preparation with gradient computation to avoid this intermediate BF16 transfer in supported backward operations. Section~\ref{sec:implementation} describes the kernel implementation.

\subsection{Weight-Gradient Accumulation and Scheduling}
\label{sec:background_accumulation}

In microbatch training, each weight-gradient GEMM adds its result to the gradients accumulated from earlier microbatches. Although the GEMM operands use FP4, the weight gradients are accumulated in FP32. Each microbatch updates the same FP32 accumulation buffer.

GPU GEMM libraries support accumulation into an existing output matrix through \(C \leftarrow \alpha AB + \beta C\). For a zero-initialized accumulator, \(\beta=0\) computes the first contribution without reading the previous output, while \(\beta=1\) adds subsequent contributions to the existing result. When these GEMMs execute as separate kernels, the intermediate FP32 accumulator is stored in global memory and accessed again by subsequent GEMMs.

Figure~\ref{fig:kernel_boundary} shows this execution in Transformer Engine~\cite{transformerengine}. The first microbatch GEMM initializes the FP32 accumulator, while subsequent GEMMs read its existing values, add their contributions, and write the updated result to global memory.

Training systems also reduce the overhead of executing multiple microbatches. Megatron-LM supports gradient accumulation and overlaps gradient communication with backward computation~\cite{NVIDIA2026MegatronAccumulation}. Pipeline scheduling methods such as GPipe, PipeDream, and Zero Bubble change the execution order of microbatches to reduce pipeline idle time~\cite{Huang2018GPipeET,Narayanan2019PipeDreamGP,Qi2024ZeroB,Qi2024PipelinePW}. These schedules can still execute weight-gradient GEMMs separately, with each GEMM accessing the FP32 accumulator in global memory.

CUTLASS supports grouped GEMMs, allowing several matrix multiplications to execute within a single kernel~\cite{NVIDIA2026CUTLASSGrouped}. Grouping reduces kernel scheduling overhead, but independent GEMMs do not automatically share an accumulator. Avoiding repeated FP32 global-memory updates requires successive GEMMs to accumulate into the same on-chip values before writing the result back.

    \section{FP4 Backward Execution: Constraints and Costs}
    
    \label{sec:formulation}

    Figures~\ref{fig:numerical_boundary} and~\ref{fig:kernel_boundary} show two sources of intermediate global-memory traffic in FP4 backward execution: BF16 gradients written before FP4 conversion and repeated FP32 weight-gradient updates across microbatches.

    We formulate the numerical and kernel constraints underlying these transfers, quantify their global-memory costs, and examine the memory required to reduce them. For a linear layer, let $X_j\in\mathbb{R}^{K\times I}$ and $G_j\in\mathbb{R}^{K\times O}$ denote the input activations and output gradients of microbatch $j$, for $j=1,\ldots,m$. The weight-gradient contribution is $X_j^\mathsf{T}G_j\in\mathbb{R}^{I\times O}$.

    \subsection{Numerical Boundary: NVFP4 Gradient Quantization}
    
    \label{sec:formulation_conversion}
    
    The numerical boundary arises when FP4 conversion requires a tensor-wide scale that is unavailable during gradient computation.   Let $S_j$ denote the FP32 global scaling factor used to quantize gradient $G_j$. We denote its rowwise and columnwise FP4 representations by $Q_r(G_j;S_j)$ and $Q_c(G_j;S_j)$. Each representation contains E2M1 values and E4M3 block scaling factors arranged in the layout required by the corresponding Blackwell GEMM.

    We eliminate the tensor-wide reduction by determining the FP32 global scaling factor from earlier gradient statistics before the gradient GEMM begins. Let \(H_{t-1}\) denote the running maximum of gradient magnitudes observed before training iteration \(t\). We select the scaling anchor as \(U_t=\gamma H_{t-1}\), where \(\gamma\) controls the margin above the previously observed maximum. The FP32 global scale \(S_j\) is then determined from this anchor according to the NVFP4 scaling rule.
    The E4M3 block scaling factors are computed from the current gradient in groups of 16 values. With the global scale already available and the block scales computed locally, the kernel can generate the E2M1 values and E4M3 scaling factors without a reduction across thread blocks. 
This removes the scale dependency that would otherwise
prevent FP4 preparation within the gradient-producing kernel.
 We examine the choice of \(\gamma\) in Appendix~\ref{app:margin}.

    Fusing FP4 preparation with gradient computation must preserve the quantized operands. 
    We compare fused and separate FP4 preparation using the same FP32 global scale, E4M3 block-scaling rules, and FP4 rounding rules. The rowwise and columnwise representations must satisfy
    
    \begin{equation}
    \begin{aligned}
    Q_r^{\mathrm{fused}}(G_j;S_j)
    &= Q_r^{\mathrm{separate}}(G_j;S_j),\\
    Q_c^{\mathrm{fused}}(G_j;S_j)
    &= Q_c^{\mathrm{separate}}(G_j;S_j).
    \end{aligned}
    \label{eq:conversion_equivalence}
    \end{equation}
    
    The equality includes the E2M1 values, E4M3 block scaling factors, and their operand layouts.

    \subsection{Kernel Boundary: FP32 Accumulation}
    
    \label{sec:formulation_accumulation}
    
    The kernel boundary arises when successive weight-gradient GEMMs cannot retain the FP32 accumulator across separate kernel executions. For microbatch $j$, let $A_j$ and $B_j$ denote its FP4 operands and $C_j$ the FP32 result of the corresponding GEMM, including the E4M3 block scaling factors. Let $\alpha_j$ denote the product of the FP32 global scaling factors associated with the two operands. The reference updates the FP32 weight-gradient buffer in microbatch order:
    
    \begin{equation}
    R_j=\operatorname{fl}_{32}\left(
    R_{j-1}+\operatorname{fl}_{32}(\alpha_j C_j)
    \right),\qquad j=1,\ldots,m.
    \label{eq:accumulation}
    \end{equation}
    
    Here, $\operatorname{fl}_{32}$ denotes FP32 rounding. The recurrence specifies a rounded contribution followed by an FP32 accumulator update. Bitwise agreement requires identical FP4 operands, global scaling factors, GEMM results, and the same FP32 rounding and accumulation sequence as the reference. Combining microbatches into one larger GEMM can change the order of multiplication and accumulation, producing different FP32 results.

    Partition the microbatches into $w$ nonempty contiguous windows $\mathcal{W}=(W_1,\ldots,W_w)$. Within each window, successive FP4 GEMMs update the same FP32 accumulator tile in registers. The tile is written to global memory at the end of the window and reloaded at the beginning of the next window. The updates follow Eq.~\eqref{eq:accumulation} in the original microbatch order.

    Each FP32 accumulator tile contains $P$ values and occupies $4P$ bytes. With $m$ separate GEMMs, the accumulator is written $m$ times and read $m-1$ times, assuming it starts at zero. Grouping the GEMMs into $w$ windows reduces this to $w$ writes and $w-1$ reads. The total memory traffic is
    
    \begin{equation}
    D_{\mathrm{acc}}(\mathcal{W})=4P(2w-1).
    \label{eq:accumulator_bytes}
    \end{equation}
    
    The memory traffic saved relative to separate GEMM execution is
    
    \begin{equation}
    \Delta D_{\mathrm{acc}}=8P(m-w).
    \label{eq:accumulator_savings}
    \end{equation}

Each eliminated accumulator store and reload therefore
saves $8P$ bytes of logical global-memory traffic.
Equations~\ref{eq:accumulator_bytes} and~\ref{eq:accumulator_savings} assume that the first window
initializes the accumulator without a global-memory
read. Our implemented kernel instead loads the FP32 weight-gradient accumulator once per window, including the first, and stores the updated result once per window. Thus, its
accumulator traffic is $8Pw$ bytes, compared with
the idealized $4P(2w-1)$ bytes. All counts represent
logical accesses, not physical DRAM traffic.

    \subsection{FP4 Operand Storage}
    
    \label{sec:formulation_windows}

    Each group of 16 NVFP4 values requires 8 bytes for the E2M1 values and 1 byte for the E4M3 scaling factor. The weight-gradient GEMM uses two operands containing $KI$ and $KO$ elements, respectively. Ignoring padding, additional layouts, and outer scaling factors, their combined storage is
    
    \begin{equation}
    S_{\mathrm{pair}}=\frac{9}{16}K(I+O).
    \label{eq:operand_storage}
    \end{equation}

    Let $M_{\mathrm{ret}}(\mathcal{W})$ denote the peak memory occupied by retained FP4 operands under schedule $\mathcal{W}$, and let $M_{\mathrm{avail}}$ denote the available memory budget. Retaining more microbatches reduces intermediate FP32 accumulator accesses but increases peak operand storage.

    \subsection{Accumulation-Window Selection}
    
    \label{sec:formulation_cost}

    Longer accumulation windows do not always improve performance. Executing several GEMMs within one kernel can be slower than using separate optimized GEMM kernels, even when the number of FP32 memory accesses is reduced. The best window size therefore depends on GEMM dimensions, execution time, and available GPU memory.

    A training step contains weight-gradient GEMMs from different layers and projections. Let $\mathcal{P}$ denote the set of layer projections and $\mathcal{W}_p$ the accumulation-window partition selected for projection $p$. We formulate window selection as minimizing training-step time while preserving the FP32 accumulation results:
    
    \begin{equation}
    \label{eq:window_optimization}
    \begin{aligned}
    \min_{\{\mathcal{W}_p\}_{p\in\mathcal{P}}}\quad
    & T_{\mathrm{step}}(\{\mathcal{W}_p\}) \\
    \text{subject to}\quad
    & R_{p,m}(\mathcal{W}_p)=R_{p,m}^{\mathrm{ref}},\\
    & \hspace{1em}\forall p\in\mathcal{P}, \\
    & M_{\mathrm{ret}}(\{\mathcal{W}_p\})\leq M_{\mathrm{avail}}.
    \end{aligned}
    \end{equation}
    
    The first constraint requires bitwise agreement with the matched FP32 reference. The second limits peak retained operand storage across the training schedule. Window assignments must also respect operand availability, accumulation order, and GPU resource limits. We evaluate selected window sizes and projection combinations rather than implementing a general scheduling optimizer.

\section{GPU Implementation}

\label{sec:implementation}

To address the numerical and kernel boundaries, we implement
fused NVFP4 gradient preparation and FP32 weight-gradient
accumulation on NVIDIA Blackwell. Both kernels use native
Blackwell FP4 GEMMs.

\subsection{Fused FP4 Gradient Preparation}

\label{sec:fused_quantization}

We integrate FP4 quantization and operand preparation into the activation backward kernel, which computes the SiLU-multiply gradient. The kernel uses the computed gradient values to produce E2M1 operands and E4M3 block scaling factors in the rowwise and columnwise layouts required by subsequent GEMMs. 

\textbf{FP32 global scaling: }The kernel reads the FP32 global scaling factor established from earlier gradient statistics before the current gradient GEMM begins. It uses this factor to compute the E4M3 block scaling factors and quantize the gradient without waiting for a tensor-wide reduction. The resulting FP4 operands match those produced by separate preparation under the same scaling state and rounding rules.

\textbf{Block scaling and operand layouts: } The gradient GEMM produces output values in registers. The kernel computes the maximum absolute value of each 16-element group, determines its E4M3 block scaling factor, and encodes the gradient values in E2M1. It prepares the rowwise and columnwise FP4 operand layouts within the gradient-producing kernel, using the gradient values before they are written to global memory. Both representations are written in the layouts required by native Blackwell GEMMs, avoiding a separate read of the BF16 gradient.

\textbf{Supported backward computations: }For the gate and up projections, the fused kernel produces FP4 operands without writing an intermediate BF16 gradient. Other projections retain higher-precision gradients required by their remaining backward computations. All FP4 operands and scale metadata required by subsequent GEMMs are stored in global memory.

\subsection{FP32 Weight-Gradient Accumulation}

\label{sec:register_accumulation}

We execute successive weight-gradient GEMMs within one CUDA thread block while retaining the accumulated FP32 output tile in registers. The FP4 operands from each microbatch are stored until the corresponding accumulation window is ready. The thread block then executes the GEMMs and completes the FP32 accumulation within that kernel invocation. It does not remain active between microbatch backward computations.

\textbf{FP32 updates across microbatches: }Each microbatch GEMM uses independently quantized E2M1 operands, E4M3 block scaling factors, and FP32 global scaling factors. The kernel executes each GEMM separately rather than concatenating the operands into a larger GEMM. After each GEMM, the FP32 result is multiplied by the product of the two FP32 global scaling factors and added to the register accumulator. The multiplication and addition use separate FP32 round-to-nearest operations, following Eq.~\eqref{eq:accumulation} and matching the arithmetic sequence of the sequential reference. The contributions are accumulated in their original microbatch order, and the FP32 result is written to \texttt{main\_grad} at the end of the window.

\textbf{Kernel resources: }The characterized Blackwell kernel uses 256 threads per CUDA thread block, 242 registers per thread, and 73,792 bytes of shared memory per block. The compiler reports 8 bytes of local memory per thread. This configuration permits one resident thread block per SM. The FP32 accumulator remains in registers while the kernel stages the FP4 operands and scaling factors for each GEMM. A four-microbatch window eliminates three intermediate FP32 accumulator store-and-reload operations.

\subsection{Accumulation Windows and Training Integration}

\label{sec:implementation_windows}

We support different accumulation windows for different layer projections. With a four-microbatch window, one kernel accumulates four weight-gradient contributions before writing the FP32 result. With two windows of two microbatches, the first kernel writes the partial result, and the second reloads it before processing the remaining contributions. Both follow the same FP32 accumulation order.

\textbf{FP4 operand storage: }FP4 operands remain in GPU memory until their scheduled weight-gradient GEMMs execute. We implement four-microbatch accumulation for selected MLP projections, two-microbatch accumulation across all projections, and four-microbatch accumulation across all projections. These configurations change the number of intermediate FP32 accumulator accesses and the peak memory used to retain FP4 operands. Section~\ref{sec:evaluation} reports their memory requirements and training-step times.

\textbf{Training integration: }We integrate both kernels into the supported linear projections of Llama 3.2 3B. The fused gradient kernels generate FP4 operands for subsequent backward GEMMs. The weight-gradient kernels consume the retained operand pairs according to the accumulation windows. The resulting FP32 weight gradients are passed to the existing optimizer without changing its interface.

\textbf{CUDA Graph execution: }We capture the Transformer Engine reference and our implementation separately using CUDA Graphs to reduce host-dispatch overhead and compare complete training steps under matched configurations. Stochastic-rounding offsets advance across graph replays, and the running gradient statistics update the FP32 global scaling factors. These updates prevent replay from reusing fixed random-number streams or stale scaling factors.

\section{Evaluation}
\label{sec:evaluation}

We evaluate FP4 backward execution on NVIDIA Blackwell using native NVFP4 GEMMs and on AMD MI210 using software-emulated MXFP4 with BF16 GEMMs. The experiments examine gradient preparation, FP32 accumulation, memory traffic and training-step performance, numerical fidelity, and sustained training behavior in Appendix.

\subsection{Experimental Setup}

\label{sec:eval_setup}

\noindent\textbf{NVIDIA Blackwell:}
For continued pretraining, we use the FineWeb-Edu dataset~\cite{Penedo2024TheFD}, starting from a pretrained Llama 3.2 3B checkpoint.

We use an NVIDIA RTX PRO 6000 Blackwell Server Edition with a 94-streaming-multiprocessor (SM) Multi-Instance GPU (MIG) 2g.48gb partition, CUDA 12.8, PyTorch 2.10, and Transformer Engine 2.21. We evaluate a 28-layer Llama 3.2 3B architecture with hidden dimension 3072 and feed-forward dimension 8192. Its 196 attention and MLP projection matrices use native NVFP4 GEMMs, while the language-model head uses BF16. Unless otherwise stated, we use a sequence length of 1024 and four microbatches per optimizer step, totaling 4096 tokens. We report GPU kernel execution time, CUDA Graph training-step throughput, and sustained throughput during continued pretraining as separate measurements.

Our baseline is Transformer Engine's native NVFP4 implementation, which accumulates FP32 weight gradients in global memory. Our implementation combines fused FP4 gradient preparation with FP32 register accumulation. For numerical comparisons, we also use separate FP4 preparation with the same scaling factors, rounding rules, and operand layouts as the fused implementation.

\noindent\textbf{AMD MI210: }We use an AMD MI210 with ROCm 7.14. Since MI210 does not support native FP4 matrix multiplication, we emulate MXFP4 using E2M1 values, power-of-two scales shared by 32 elements, and BF16 GEMMs. 
Native MXFP4 training has also been investigated on newer
AMD MI355X hardware~\cite{Cim2026PretrainingLL}. We compare three implementations using the same prepared FP4 operands: (1) optimized library BF16 GEMMs that update the FP32 accumulator in global memory, (2) custom BF16 GEMMs that execute separately and update the FP32 accumulator in global memory, and (3) custom BF16 GEMMs that execute within one kernel and retain the FP32 accumulator in registers across microbatches. We additionally examine fused and separate MXFP4 gradient preparation under the same block-scaling and rounding rules.

\begin{figure*}[t]
    \centering
    \includegraphics[
        width=0.40\textwidth
    ]{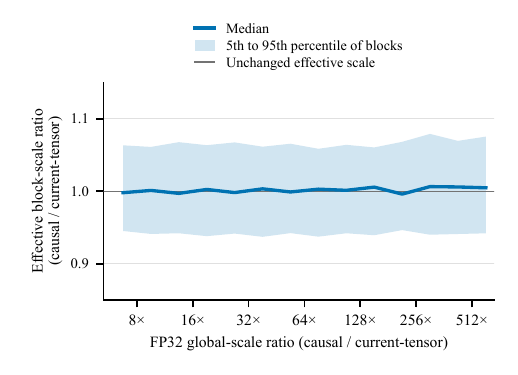}%
    \hspace{0.015\textwidth}%
    \includegraphics[
        width=0.40\textwidth,
    ]{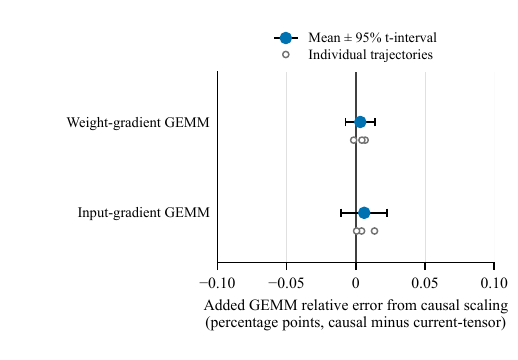}
    \includegraphics[
        width=0.40\textwidth
    ]{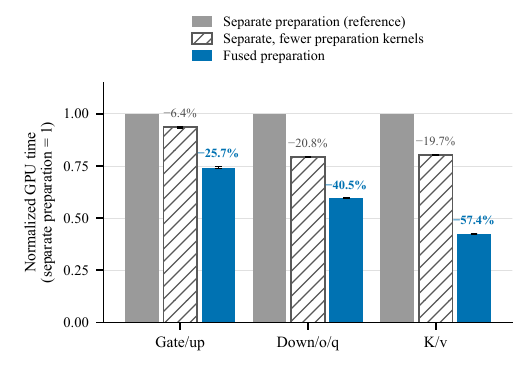}%
    \hspace{0.015\textwidth}%
    \includegraphics[
        width=0.40\textwidth
    ]{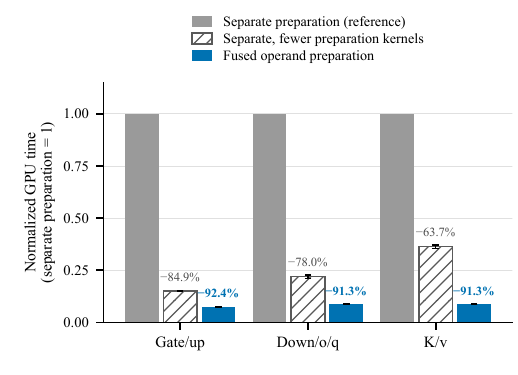}

    \caption{\textbf{Numerical boundary evaluation.}
    (a) NVFP4 scale compensation.
    (b) Change in backward-GEMM error.
    (c) Blackwell NVFP4 preparation.
    (d) MI210 MXFP4 preparation.}
    \label{fig:numerical_boundary_eval}
\end{figure*}

\subsection{Numerical Boundary}
\label{sec:numerical_boundary_eval}

Figure~\ref{fig:numerical_boundary_eval}(a) shows that
effective block scales remain close to their current-tensor
values even when causal and current-tensor outer scales
differ substantially. This confirms the local compensation
behavior described in Section~\ref{sec:formulation_conversion}.
The plotted distribution covers blocks with normal,
unsaturated E4M3 scales under both policies; the appendix
characterizes blocks outside this range.
Figure~\ref{fig:numerical_boundary_eval}(b) shows that the
corresponding changes in native FP4 backward-GEMM error
remain close to zero across the evaluated training
trajectories. Thus, the numerical benefit of current-block
adaptation persists despite using an earlier outer scale,
with the measured residual including the complete
quantization and GEMM path.

On Blackwell, Fig.~\ref{fig:numerical_boundary_eval}(c)
shows lower preparation time across all three projection
families. For gate/up, producer-local preparation also
avoids materializing the intermediate BF16 gradient.
For down/o/q and k/v, the gains come from combining
preparation operations after gradient materialization.
The hatched bars show the effect of removing the amax
pass as a timing-only control. At sequence length 2048, four-microbatch accumulation across all projections exhausts the available GPU memory on the tested Blackwell MIG partition, while two-microbatch windows across all projections remain executable. This illustrates how shorter accumulation windows allow broader projection coverage under memory constraints.

On AMD MI210, Fig.~\ref{fig:numerical_boundary_eval}(d)
shows that combining MXFP4 operand-preparation operations
also reduces execution time across all three families.
These measurements begin with materialized BF16 gradients
and use emulated MXFP4. Separately, our producer-fused
gate/up implementation eliminates this intermediate
materialization while preserving the matched MXFP4
operands under identical stochastic rounding.

\subsection{Kernel Boundary: FP32 Accumulation}
\label{sec:accumulation_eval}

\begin{figure}[t]
    \centering
    \includegraphics[width=0.8\columnwidth]
        {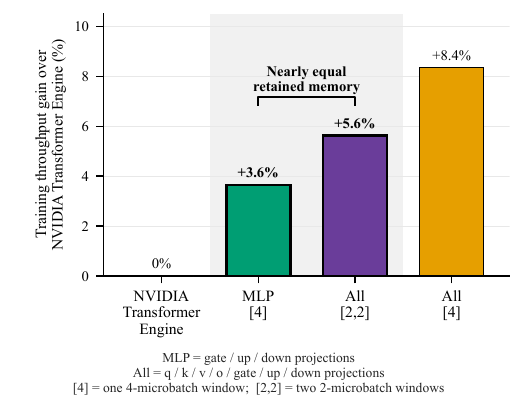}

\caption{\textbf{FP32 accumulation on Blackwell.}
Training throughput across accumulation-window configurations.}
\label{fig:accumulation_window}

\end{figure}

\begin{figure*}[t]
    \centering

    \includegraphics[width=0.40\textwidth]
        {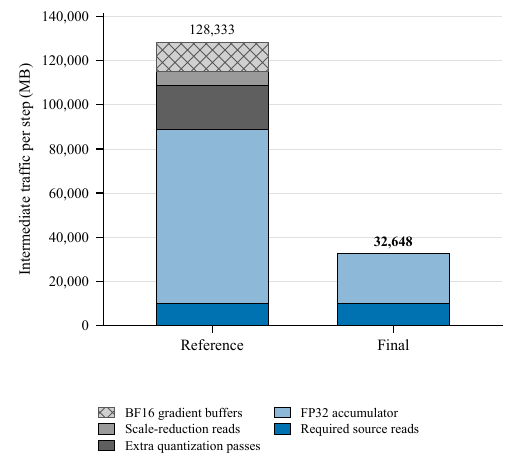}
    \hfill
    \includegraphics[
    width=0.4\textwidth]{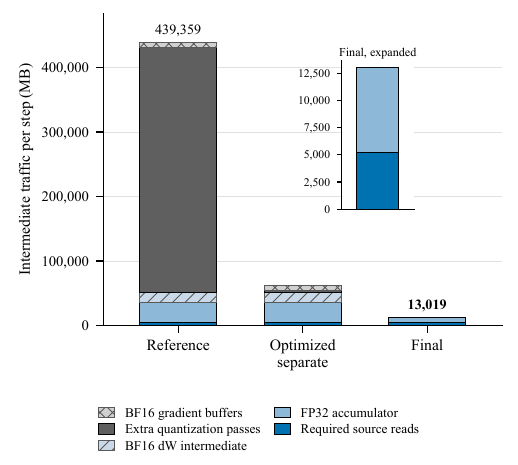}

    \caption{\textbf{Intermediate global-memory traffic.}
   Calculated intermediate global-memory reads and writes per training step for reference and optimized FP4 backward execution on
    (a) NVIDIA Blackwell and (b) AMD MI210.}
    \label{fig:traffic_composition}

\end{figure*}

\begin{figure*}[t]
    \centering
    \begin{minipage}[c]{0.40\textwidth}
        \centering
        \includegraphics[width=\linewidth]
            {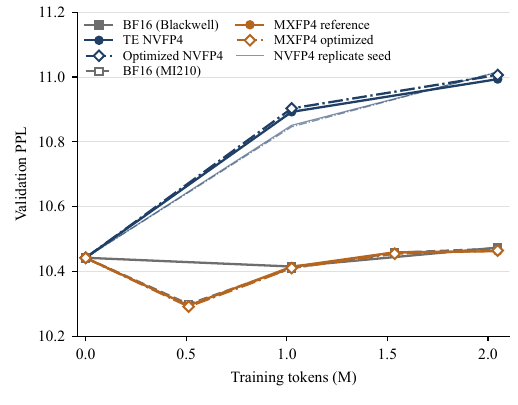}

        \small (a) Validation perplexity
    \end{minipage}
    \hfill
    \begin{minipage}[c]{0.25\textwidth}
        \centering
        \includegraphics[width=\linewidth]
            {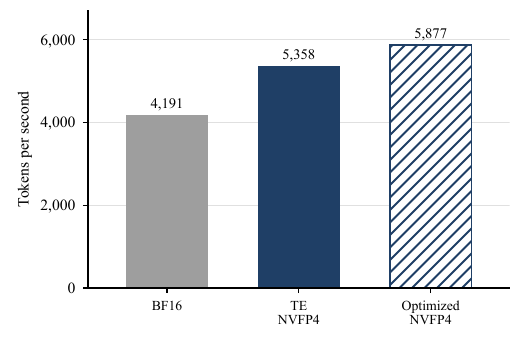}

        \small (b) Blackwell throughput
    \end{minipage}
    \hfill
    \begin{minipage}[c]{0.25\textwidth}
        \centering
        \includegraphics[width=\linewidth]
            {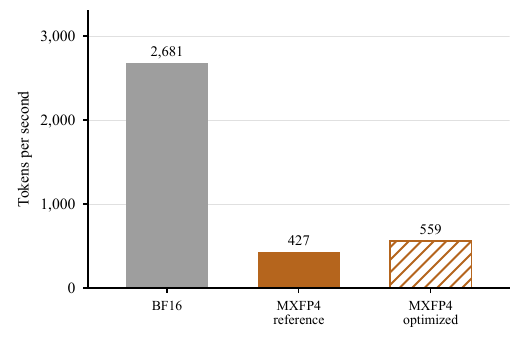}

        \small (c) MI210 throughput
    \end{minipage}

    \caption{\textbf{End-to-end FP4 training.}
    Validation perplexity over two million training
    tokens and measured training throughput for native
    NVFP4 and emulated MXFP4 execution.}
    \label{fig:end_to_end}

\end{figure*}

Figure~\ref{fig:accumulation_window} shows that
accumulating across all projections gives the largest
training-throughput improvement with a full window.
At nearly equal additional retained-operand memory (841 MiB), two-microbatch windows across all projections achieve shorter training-step time than four-microbatch accumulation restricted to MLP projections. Thus,
covering more projections can be more effective than
extending the accumulation window for fewer projections.
When our GEMMs execute separately, they are slower than
the optimized library GEMMs, confirming that the benefit
does not come from faster individual GEMMs. The register-resident implementation preserves the sequential FP32 accumulation result bit for bit across all tested weight-gradient tensors. Two-microbatch windows across all projections recover 69\% of the training-step time reduction achieved by four-microbatch accumulation across all projections, while using roughly one-third of the additional retained-operand memory.

On AMD MI210, the same window ordering appears in the
weight-gradient computation using emulated MXFP4 and
BF16 GEMMs. At nearly equal retained memory, shorter
windows across all projections outperform full-window
accumulation restricted to MLP projections. However,
the optimized library GEMMs remain faster overall,
and register-resident accumulation produces no
measurable whole-step improvement. Unlike Blackwell,
the AMD implementation retains BF16 operands rather
than FP4 operands. The AMD implementation details and numerical validation are provided in Appendix~\ref{app:formats} and~\ref{app:implementation}.

\subsection{Memory Boundary: Intermediate Global-Memory Traffic}
\label{sec:memory_boundary}

We examine the intermediate global-memory traffic
associated with FP4 quantization and FP32 accumulation.
Figure~\ref{fig:traffic_composition} reports calculated
logical reads and writes per training step, rather than
measured HBM transactions. The evaluated configurations
accumulate gradients across four microbatches.

On Blackwell, repeated FP32 accumulator reads and
writes account for the largest share of intermediate
traffic in the reference. Register-resident accumulation
removes intermediate accumulator transfers within
each window. Producer-local quantization also removes
the separate scale-reduction reads and avoids BF16
gradient materialization for gate/up. Causal scaling
makes this possible by providing the outer scale
before gradient production; changing the scale alone
does not remove memory accesses.

For the evaluated four-microbatch window, the reference
performs four accumulator stores and three reads, while
our kernel performs one initial load and one final store.

On AMD MI210, additional MXFP4 quantization passes
dominate the reference traffic. The optimized separate
implementation removes most of these passes without
fusing gradient production. The final implementation
further reduces intermediate BF16 traffic and FP32
accumulator transfers. The two platforms therefore
benefit from removing different sources of intermediate
traffic, reflecting their native NVFP4 and emulated
MXFP4 execution paths. The optimized-separate comparison removes repeated quantization passes before evaluating the additional traffic reduction from fused execution.

Although the reference implementations have different sources of intermediate traffic, both retain substantial FP32 accumulation traffic after reducing the extra quantization passes. Keeping the accumulator in registers removes repeated transfers across microbatches, while producer-local quantization removes the supported BF16 gradient intermediates.

\subsection{End-to-End Training Performance}
\label{sec:end_to_end}

Figure~\ref{fig:end_to_end} reports validation perplexity over two million training tokens and training throughput on NVIDIA Blackwell and AMD MI210. Within each FP4 format, optimized execution closely tracks its corresponding reference under matched training and validation protocols (Figure~\ref{fig:end_to_end}(a)). The comparison evaluates execution optimizations within each format rather than numerical differences between NVFP4 and MXFP4. Longer-horizon NVFP4 results over 20 million tokens, including independent training replicas, are reported in Appendix~\ref{app:sustained_training}.

On Blackwell, producer-local FP4 preparation and register-resident FP32 accumulation reduce GPU kernel time by 9.28\% and improve CUDA Graph training throughput by 9.63\% over NVIDIA Transformer Engine NVFP4 (Figure~\ref{fig:end_to_end}(b)). On AMD MI210, optimized software-emulated MXFP4 preparation improves training throughput by 30.9\% over the corresponding MXFP4 reference (Figure~\ref{fig:end_to_end}(c)). These results demonstrate the benefits of reorganizing FP4 backward execution on native NVFP4 Tensor Cores and software-emulated MXFP4 using BF16 GEMMs, respectively.

\section{Limitations}
\label{sec:limitations}

\textbf{Training-relevant FP4 numerical error:}
Our work optimizes FP4 execution rather than the numerical differences between FP4 and BF16 training~\cite{Agrusa2026WhatMF}. Although optimized NVFP4 closely tracks Transformer Engine NVFP4, both differ from BF16 in validation behavior. In our controlled experiments, forward weight and activation quantization affect training gradients more strongly than backward quantization, motivating further investigation beyond local reconstruction error.

\textbf{Causal NVFP4 scaling:}
Historical gradient statistics make the FP32 global scale available before gradient computation but cannot guarantee coverage of future gradients. Large values may exceed the selected range, while overly conservative scaling can drive E4M3 block scales into the subnormal regime. Our experiments exhibit stable training behavior, but unconditional range coverage requires additional mechanisms, such as adaptive scaling or fallback.

\section{Conclusion}
\label{sec:conclusion}

FP4 scaling and ordered FP32 accumulation constrain arithmetic but do not inherently require intermediate global-memory transfers. We address these transfers through producer-local FP4 quantization and register-resident FP32 accumulation, achieving 9.63\% higher CUDA Graph training throughput on NVIDIA Blackwell and sustaining a 9.7\% improvement over 20 million tokens with comparable validation perplexity. Accumulation-window selection further balances retained-operand memory against projection coverage, allowing shorter windows to outperform longer windows under comparable memory budgets. On AMD MI210, optimized software-emulated MXFP4 preparation improves training throughput by 30.9\%, while register-resident accumulation remains dependent on GEMM efficiency. These results demonstrate that FP4 numerical requirements need not dictate kernel boundaries or intermediate global-memory traffic.

\bibliographystyle{IEEEtran}
\bibliography{references}

\clearpage

\appendices

\section{Numerical Requirements and Correctness}
\label{appa}
\subsection{Scale Availability in MXFP4 and NVFP4}
\label{app:formats}

MXFP4 and NVFP4 both represent individual values using E2M1, but differ in how their scales are computed. MXFP4 uses a power-of-two E8M0 scale for each block of 32 values. Since the scale depends only on the values within that block, it can be computed as soon as the block is available. NVFP4 uses an E4M3 scale for each block of 16 values, along with an FP32 global scale shared across the tensor. Transformer Engine computes this FP32 global scale from the current tensor maximum, which is not available until the complete gradient has been computed. Thus, MXFP4 allows gradient conversion using block-local information, whereas current-tensor NVFP4 scaling requires a tensor-wide maximum. Our running-maximum method makes the FP32 global scale available during gradient computation without requiring the current tensor maximum.

On AMD MI210, we implement MXFP4 conversion in software and reconstruct the quantized operands as BF16 values for matrix multiplication. We also fuse MXFP4 preparation into supported gradient-computation kernels. Unlike Blackwell, which executes native NVFP4 matrix multiplications, the AMD implementation uses BF16 GEMMs on the reconstructed operands. The two implementations therefore differ in their conversion and matrix-multiplication paths, while allowing gradient preparation to proceed without waiting for a current-tensor maximum.

We compared the numerical accuracy of MXFP4 and NVFP4 using BF16 operands captured from Llama-3.2-3B training. The sample covers five transformer layers and all seven linear projection types, including weights, activations, and backward gradients. Both formats use E2M1 values with round-to-nearest-even, without stochastic rounding or Hadamard transforms. The MXFP4 configuration uses the study's power-of-two block-scale selection rule, while NVFP4 uses E4M3 block scales and an FP32 global scale. We additionally evaluate FP32 block scales with the same E2M1 values as a high-precision scaling control. Reconstruction error is measured against the original BF16 tensors, and GEMM error against FP32 matrix multiplication on those tensors. All numerical comparisons use decoded operands and FP32 arithmetic, independently of the GPU execution measurements.

\begin{figure*}[t]
    \centering
    \includegraphics[width=0.8\textwidth]{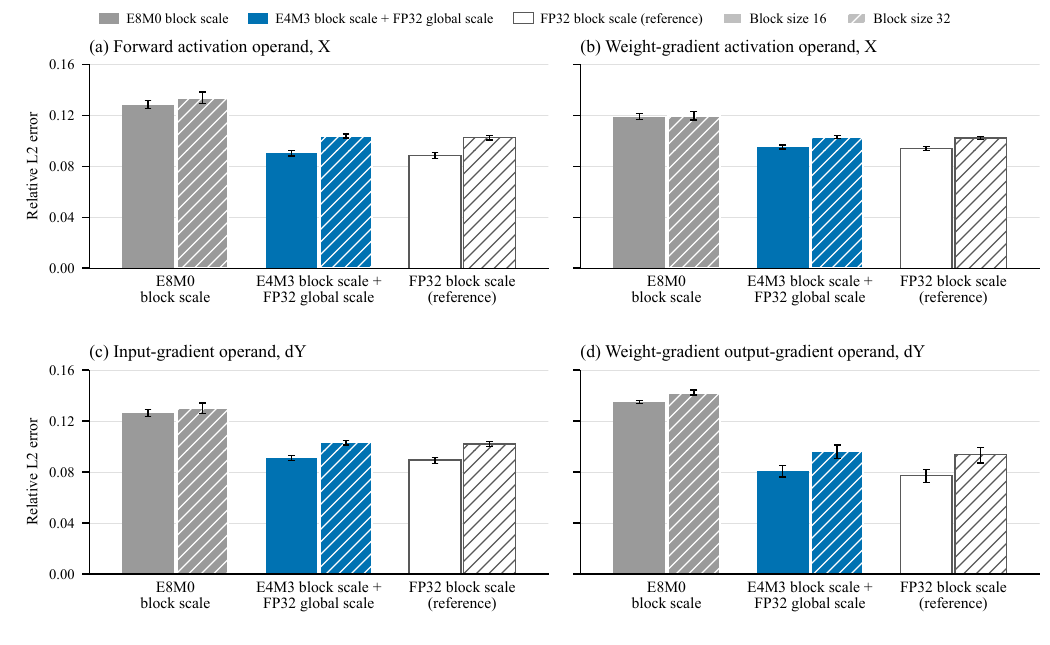}
    \caption{\textbf{FP4 operand reconstruction error.}
    Relative L2 error for activation ($X$) and
    output-gradient ($dY$) operands in forward,
    input-gradient, and weight-gradient computations.
    We compare E8M0 block scales, E4M3 block scales
    with an FP32 global scale, and FP32 block scales
    at block sizes 16 and 32. Error bars show the
    uncertainty intervals over the evaluated tensors.}
    \label{fig:fp4_block_scale_accuracy}
\end{figure*}

The scale-selection method has a greater effect on reconstruction error than the block size in the tested configurations. Fig.~\ref{fig:fp4_block_scale_accuracy} shows that reducing the block size from 32 to 16 changes the error relatively little under E8M0 scaling. With E4M3 scaling, smaller blocks generally provide a more noticeable reduction. The E4M3 results also remain close to those obtained using FP32 block scales. The larger error under E8M0 reflects both its power-of-two scale representation and the scale-selection rule used in this experiment. The results show that scale representation and selection affect reconstruction error substantially, alongside the number of values sharing each scale.

\begin{table}[t]
    \centering
    \caption{Relative L2 GEMM output error for MXFP4 and NVFP4 using identical captured BF16 operands and FP32 reference arithmetic.}
    \label{tab:fp4_format_gemm_error}
    \begin{tabular}{lcc}
        \toprule
        Computation & MXFP4 & NVFP4 \\
        \midrule
        Forward & 0.137 & 0.094 \\
        Input gradient & 0.173 & 0.133 \\
        Weight gradient (4 microbatches) & 0.154 & 0.101 \\
        \bottomrule
    \end{tabular}
\end{table}

The difference also carries into matrix multiplication. Table~\ref{tab:fp4_format_gemm_error} shows lower output error for NVFP4 in the forward, input-gradient, and weight-gradient computations. However, NVFP4 requires the FP32 global scale to be available before gradient conversion. The following subsection examines how selecting this scale from earlier gradient statistics affects native NVFP4 accuracy.

\subsection{Global Scale Selection and Numerical Accuracy}
\label{app:margin}

Section~\ref{sec:numerical_boundary_eval} shows that local E4M3 block scales compensate
for changes in the FP32 global scale within their
representable range. Here, we examine how E4M3 rounding,
saturation, and subnormal scales limit this compensation
as the scaling margin changes. Let $a_b$ denote the
maximum magnitude in block $b$, $H_{t-1}$ the running
maximum, and $\gamma$ the scaling margin. The FP32 global
decoding scale $S$ and the ideal local scale $s_b^*$ satisfy
\begin{equation}
S = \frac{\gamma H_{t-1}}{2688},
\qquad
s_b^* = \frac{448a_b}{\gamma H_{t-1}}.
\end{equation}
Their product is
\begin{equation}
S s_b^* = \frac{a_b}{6}.
\label{eq:margin_cancellation}
\end{equation}

The ideal effective block scale is therefore independent
of the global scaling margin. In practice, E4M3 rounding
changes this relationship, while saturation and subnormal
scaling introduce additional effects near the representable
limits. The resulting FP4 values need not remain identical
under different global scales.

\begin{figure*}[t]
    \centering
    \includegraphics[width=0.8\textwidth]
        {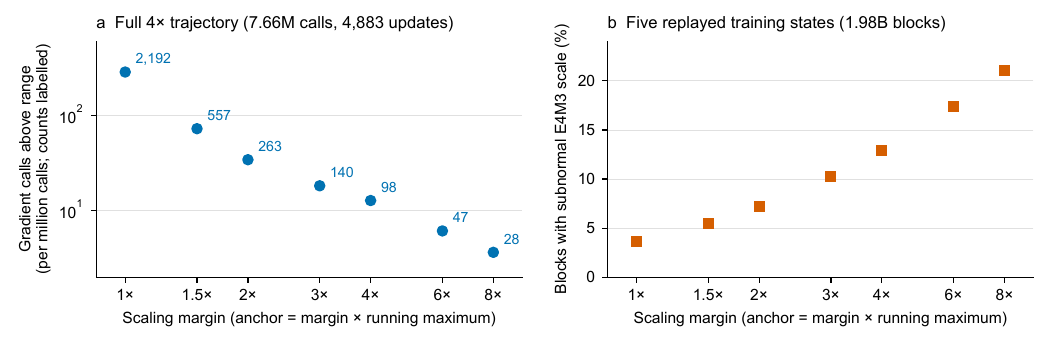}
    \caption{\textbf{Scaling margin and E4M3 representability.}
    (a) Gradient calls exceeding the range selected by
    the causal FP32 global scale, measured over 7.66
    million calls and 4,883 training updates.
    (b) Fraction of E4M3 block scales in the subnormal
    range, measured over 1.98 billion blocks from five
    replayed training states. The dashed line marks
    the deployed $4\times$ margin.}
    \label{fig:margin_range_tradeoff}
\end{figure*}

Fig.~\ref{fig:margin_range_tradeoff} shows the tradeoff
between gradient-range coverage and E4M3 block-scale
representability. Increasing the scaling margin reduces
gradient calls exceeding the selected range, from 2,192
per million calls at $\gamma=1$ to 28 at $\gamma=8$.
However, the fraction of subnormal E4M3 block scales
increases with the margin. Subnormal scales remain
representable, although their absolute spacing is fixed
and sufficiently small values can round to zero.
Increasing the margin therefore improves range coverage
while reducing the precision available to small block
scales.

\begin{figure*}[t]
    \centering
    \includegraphics[width=0.8\textwidth]
        {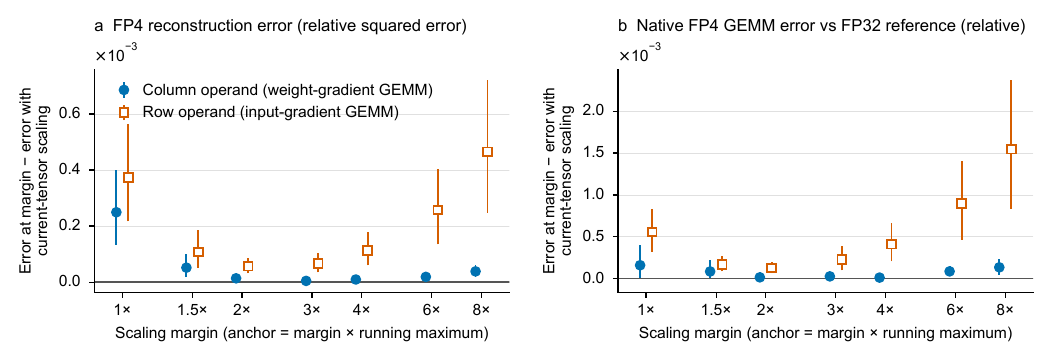}
    \caption{\textbf{Numerical error from causal FP32
    global-scale selection.}
    Additional (a) FP4 reconstruction error and
    (b) native FP4 GEMM output error relative to
    current-tensor scaling on matched gradients.
    Columnwise weight-gradient and rowwise input-gradient
    operands are shown separately. Error bars retain
    the clustered uncertainty intervals from the
    original analysis. The dashed line marks
    $\gamma=4$.}
    \label{fig:margin_numerical_error}
\end{figure*}

Fig.~\ref{fig:margin_numerical_error} compares
running-maximum scaling with current-tensor scaling
on identical gradients. The additional reconstruction
and GEMM errors are smallest across intermediate
margins and increase toward the extremes. This
behavior is consistent with insufficient range at
smaller margins and reduced precision for small
block scales at larger margins. Input-gradient
computations are more sensitive to larger margins
than weight-gradient computations. Although many
E4M3 scales become subnormal, they can still represent
nonzero values. Subnormal-scale incidence therefore
does not directly measure the amount of gradient
information lost.

\begin{figure*}[t]
    \centering
    \includegraphics[width=0.8\textwidth]
        {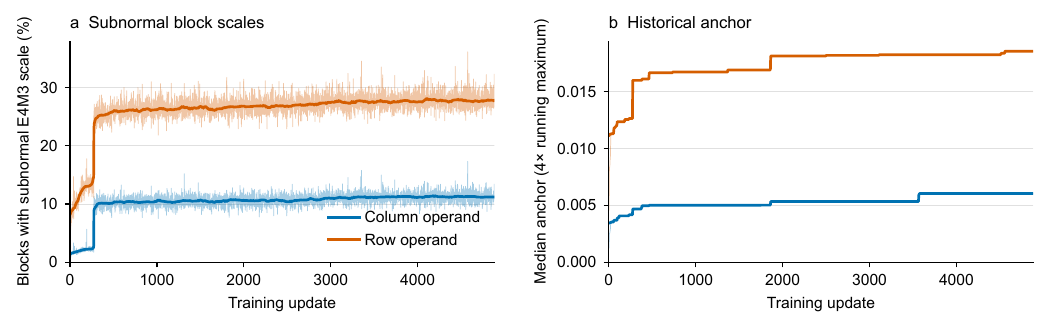}
    \caption{\textbf{Causal scaling behavior during training.}
    (a) Fraction of subnormal E4M3 block scales for
    rowwise and columnwise operands, including per-update
    measurements and 101-update rolling medians.
    (b) Historical gradient-scale anchors selected from
    running maxima. Measurements use the deployed
    $\gamma=4$ margin.}
    \label{fig:margin_trajectory}
\end{figure*}

The running maximum does not decrease when subsequent
gradients become smaller. Fig.~\ref{fig:margin_trajectory}
shows that the historical anchor increases in steps
throughout training. Subnormal-scale incidence rises
early and subsequently changes more gradually, with
a larger fraction for rowwise than columnwise operands.
A windowed or decaying maximum could limit the influence
of earlier large gradients without changing the
requirement that the FP32 global scale be available
before gradient computation. The numerical behavior
of such a policy is not evaluated here.

\begin{figure}[h]
    \centering
    \includegraphics[width=\linewidth]
        {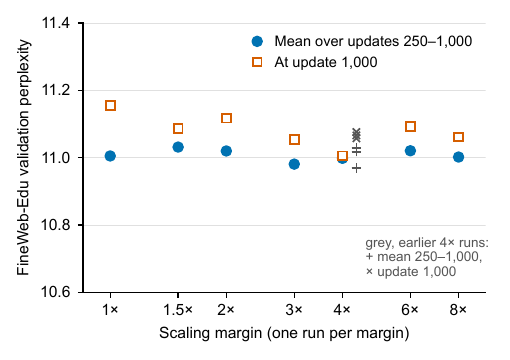}
    \caption{\textbf{Validation perplexity across causal
    scaling margins.}
    FineWeb-Edu validation perplexity is reported as
    the mean over updates 250--1,000 and at update 1,000.
    Each margin has one training run; independent earlier
    runs at $\gamma=4$ are shown separately. The dashed
    line indicates the deployed margin.}
    \label{fig:margin_validation}
\end{figure}

Fig.~\ref{fig:margin_validation} shows validation
perplexity across the evaluated scaling margins.
The results remain within a relatively narrow range,
although differences between individual runs and
evaluation checkpoints are visible. Since each margin
has one training run, these measurements do not
establish a statistically optimal margin.

We select $\gamma=4$ to balance gradient-range
violations and input-gradient error. This margin
reduces range violations compared with smaller
margins while avoiding the greater input-gradient
error observed at larger margins. We use $\gamma=4$
throughout our Blackwell experiments. The AMD MXFP4
implementation uses block-local power-of-two scales
and does not require this FP32 global scaling margin.

These results characterize the numerical cost of
selecting the FP32 global scale before gradient
computation. Local E4M3 scales compensate for much
of the change in the global scale, while their finite
representation determines the residual reconstruction
and GEMM error.

\section{GPU Kernel Implementation}
\label{app:implementation}

For a linear layer with input $X$, weights $W$, and
output gradient $dY$, the forward and backward
matrix multiplications are
\begin{equation}
Y = XW^\top,\qquad
dX = dY W,\qquad
dW = dY^\top X.
\end{equation}
The activation gradient $dX$ produced by one layer
becomes an operand for the preceding layer's backward
computation. FP4 GEMMs require operands in layouts
determined by their use in matrix multiplication.
For a newly computed gradient $G$, we prepare
\begin{equation}
(Q_r,S_r)=\mathcal{Q}_r(G),\qquad
(Q_c,S_c)=\mathcal{Q}_c(G),
\label{eq:dual_orientation}
\end{equation}
where $Q_r$ and $Q_c$ contain E2M1 values in the
required rowwise and columnwise layouts, and $S_r$
and $S_c$ contain their respective block scales.
The two conversions use the same source values,
but their block scales and quantized values can
differ because the block boundaries differ.

\subsection{NVIDIA Blackwell Implementation}
\label{app:nvidia_impl}

Our Blackwell implementation builds on NVIDIA
Transformer Engine~\cite{transformerengine} and uses
native NVFP4 Tensor Core GEMMs. NVFP4 uses E4M3
block scales for groups of 16 values and an FP32
global scale. The reference prepares the required
FP4 operand layouts in separate conversion kernels
after the BF16 gradient has been materialized.
Our implementation prepares both orientations
from the gradient values produced within the
supported gate/up activation-backward kernels.
The two conversions share the source gradient
values while computing their own E4M3 block
scales and E2M1 outputs. The FP32 global scale
comes from earlier gradient statistics, removing
the need to wait for the current gradient maximum.

For weight-gradient accumulation, the reference
updates an FP32 gradient buffer across separately
executed microbatch GEMMs. Our kernel instead
executes the microbatch GEMMs in order while
retaining each output tile's FP32 accumulator in
registers. The kernel is launched after the
operands for its accumulation window are available.
The operands retain their independently computed
FP4 scales, and accumulation follows the original
microbatch order.

\begin{figure}[t]
    \centering
    \includegraphics[width=0.8\linewidth]
        {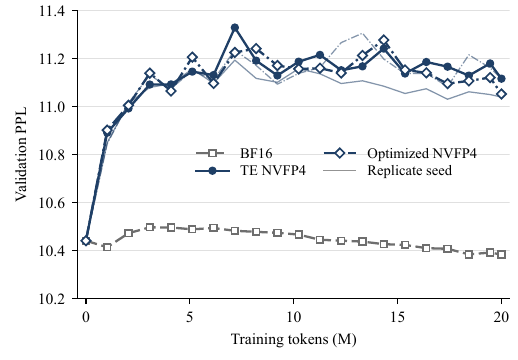}
    \caption{\textbf{Validation perplexity over 20 million
    training tokens on NVIDIA Blackwell.}
    BF16, Transformer Engine NVFP4, and optimized NVFP4
    are evaluated under the same training configuration.
    The additional NVFP4 trajectory shows an independent
    run with a different seed.}
    \label{fig:nvidia_convergence}
\end{figure}

\subsection{AMD MI210 Implementation}
\label{app:amd_impl}

Our AMD implementation uses MXFP4 quantization,
with E2M1 values and a power-of-two E8M0 scale
for each group of 32 values. Unlike NVFP4,
MXFP4 does not require an FP32 global scale.
We combine supported gradient computations
with the preparation of both operand orientations
in the same kernel. The computed gradient values
are reused through local data staging, while
the rowwise and columnwise outputs retain their
own block scales and layouts. On MI210, this
local storage is implemented using LDS.

MI210 does not execute native FP4 matrix
multiplication. The MXFP4 operands are therefore
reconstructed in BF16 before GEMM execution.
Our weight-gradient implementation compares
optimized library BF16 GEMMs with custom kernels
that execute successive microbatch GEMMs while
retaining the FP32 accumulator in registers.
The custom kernels preserve microbatch order
and FP32 accumulation. Unlike Blackwell, the
deferred GEMM operands on AMD are retained in BF16
rather than native FP4 format.

\section{Sustained Blackwell Training}
\label{app:sustained_training}

We evaluate sustained training over 20 million tokens
on NVIDIA Blackwell using the same model and training
configuration for BF16, Transformer Engine NVFP4,
and our optimized NVFP4 implementation. The optimized
implementation achieves 9.7\% higher sustained training
throughput than the Transformer Engine NVFP4 reference.
This result is consistent with the end-to-end
performance improvement reported in the main paper.

Fig.~\ref{fig:nvidia_convergence} shows that optimized
NVFP4 follows the Transformer Engine NVFP4 validation
trajectory over the measured training horizon. At
20 million tokens, validation perplexity is 11.053
for optimized NVFP4 and 11.117 for the Transformer
Engine reference. Both NVFP4 configurations exhibit
higher validation perplexity than BF16. These results
show that the throughput improvement is sustained
without a substantial change in validation behavior
relative to the NVFP4 reference.

\end{document}